\documentclass[aps,prb,preprint,superscriptaddress,showpacs,preprintnumbers,amsmath,amssymb]{revtex4}
\usepackage{graphicx}
\usepackage{longtable}
\usepackage{bm}
\usepackage{dcolumn}

\begin{document}

\title{Magnetoelastic and thermal effects in the BiMn$_{2}$O$_{5}$ lattice: a high-resolution x-ray diffraction study}

\author{E. Granado}
\email{egranado@ifi.unicamp.br}
\affiliation{Instituto de F\'{i}sica ``Gleb Wataghin,'' UNICAMP, CP 6165, 13083-970, Campinas, SP, Brazil}
\affiliation{Laborat\'{o}rio Nacional de
Luz S\'{i}ncrotron, C.P. 6192, 13084-971, Campinas, SP, Brazil}

\author{M. S. Eleot\'{e}rio}
\affiliation{Instituto de F\'{i}sica ``Gleb Wataghin,'' UNICAMP, CP 6165, 13083-970, Campinas, SP, Brazil}
\affiliation{Laborat\'{o}rio Nacional de
Luz S\'{i}ncrotron, C.P. 6192, 13084-971, Campinas, SP, Brazil}

\author{A. F. Garc\'{i}a-Flores}
\affiliation{Instituto de F\'{i}sica ``Gleb Wataghin,'' UNICAMP, CP 6165, 13083-970, Campinas, SP, Brazil}

\author{J. A. Souza}
\affiliation{Instituto de F\'{i}sica, Universidade de S\~{a}o Paulo, CP 66318, 05315-970, S\~{a}o Paulo, SP, Brazil}

\author{E. I. Golovenchits}
\affiliation{Ioffe Physical-Technical Institute of RAS, 194021, St. Petersburg, Russia}

\author{V. A. Sanina}
\affiliation{Ioffe Physical-Technical Institute of RAS, 194021, St. Petersburg, Russia}

\begin{abstract}

High-resolution synchrotron x-ray diffraction measurements were performed on single crystalline and powder samples 
of BiMn$_{2}$O$_{5}$. A linear temperature dependence of the unit cell volume was found
between $T_{N}=38$ K and 100 K, suggesting that a low-energy lattice excitation
may be responsible for the lattice expansion in this temperature range. Between $T^{*} \sim 65$ K
and $T_{N}$, all lattice parameters showed incipient magnetoelastic
effects, due to short-range spin correlations. An anisotropic strain along the {\bf a}-direction was also observed below $T^{*}$.
Below $T_{N}$, a relatively large contraction of the $a$-parameter following the square of the
average sublattice magnetization of Mn was found, indicating that a second-order spin hamiltonian accounts for the magnetic
interactions along this direction. On the other hand,
the more complex behaviors found for $b$ and $c$ suggest additional magnetic transitions
below $T_{N}$ and perhaps higher-order terms in the spin hamiltonian. Polycrystalline samples grown by distinct
routes and with nearly homogeneous crystal structure above $T_{N}$
presented structural phase coexistence below $T_{N}$, indicating a close competition amongst distinct magnetostructural states 
in this compound.

\end{abstract}


\maketitle

\section{Introduction}

Multiferroic materials with coexisting (anti)ferromagnetism and ferroelectricity
have attracted renewed attention, due to the interesting physics involved as well as relevant
potential applications in spintronics. A fairly strong coupling amongst ferroelectric and magnetic order parameters may occur as a result of
exchange striction effects in magnetic structures lacking an inversion center (for a recent review, see Ref. \cite{Cheong}).
Examples can be found within the class of frustrated antiferromagnets.
The $R$Mn$_{2}$O$_{5}$ family is a particularly interesting case,\cite{Hur1,Hur2} in which
Mn$^{4+}$O$_{6}$ octahedra and Mn$^{3+}$O$_{5}$ pyramids are
interconnected and no possible spin configuration can simultaneously satisfy all nearest-neighbor
Mn-O-Mn superexchange interactions.\cite{Chapon,Blake,Chapon2}
As a consequence, the magnetic structures actually found in
this family frustrates some of the spin interactions. Below the magnetic ordering temperature,
slight atomic displacements take place and strengthen (weaken) the satisfied (frutrated) interactions,
breaking the inversion symmetry of the structure and leading to ferroelectricity.\cite{Chapon,Blake,Chapon2,Wang}

While the above mechanism explains qualitatively the multiferroic behavior of $R$Mn$_{2}$O$_{5}$,
detailed experimental information on the atomic displacements associated with each of the commensurate or incommensurate
spin structures of this family is still lacking. This is mostly due to the small magnitude of such displacements,
presumably below $\sim 0.01$ \AA.\cite{Wang} This limitation prevents a more quantitative test for the existing theories, most
noticeably for ab-initio calculations.\cite{Wang}
On the other hand, the lattice parameters can be obtained directly by high-resolution x-ray
diffraction experiments, and the thermal expansion coefficients may be also accurately obtained by macroscopic dilatometry
measurements on single crystals. Such measurements
may reveal the overall magnetoelastic coupling in the unit cell dimensions,
and carry relevant information on the microscopic spin-lattice coupling mechanism leading to ferroelectricity.
Dilatometry measurements of thermal expansion coefficients have been carried out for $R=$ Ho, Dy, and Tb,\cite{delaCruz}
clearly revealing the lattice anomalies related to each of the magnetic transitions of these materials.

It is well known that thermal expansion coefficients obtained by macroscopic dilatometry may have much
higher resolution than the typical results of x-ray or neutron diffraction. On the other hand, diffraction
is the only choice to investigate powder samples with anisotropic crystal structure, and may be also useful for single
crystal studies if simultaneous lattice expansion and strain broadening measurements are required. This is generally the case
in investigations of ferroelectric materials,\cite{Doriguetto} which commonly have distinct behaviors for powder and single crystalline
samples, also depending on details of sample growth. Thus, a technique combining the attributes of high resolution to detect
minute lattice anomalies and microscopic sensitivity to probe powder samples and/or inhomogeneous phases is highly desirable
to study ferroelectric materials, in particular the multiferroics. In fact, high-resolution
synchrotron x-ray diffraction may be the technique of choice for some of these cases.

BiMn$_{2}$O$_{5}$ presents a magnetic structure with propagation vector $\tau=(1/2,0,1/2)$ at low
temperatures,\cite{Bertaut,Munoz} while other members show ground states with
$\tau=(k_{x},0,k_{z})$ with $k_{x} \sim 1/2$ and $0.25 \leq k_{z} \leq 0.37$.\cite{Buisson,Gardner,Wilkinson,Chapon,Blake,Chapon2}
Structurally, the major differences to the other compounds of the series arise from a largely distorted BiO$_{8}$ cage,
which has been ascribed to the electron lone pair in Bi$^{3+}$.\cite{Munoz} No detailed study on the magnetoelastic properties
of this compound has been performed, to our knowledge.
To bridge this gap, we performed a synchrotron x-ray diffraction study on a single crystal of BiMn$_{2}$O$_{5}$, as well
as on two powder samples grown by distinct routes. It is shown that this technique may have enough resolution to reveal
the relatively subtle lattice parameter anomalies related to the magnetic transitions and may provide
quantitative details of the spin-lattice coupling in this compound.

\section{Experimental Details}

The single crystal used in the present study were prepared by the flux method, as described elsewhere.\cite{Hur2,Sanina} 
In addition, two polycrystalline samples of BMO were grown by entirely different routes. The sample named
BMO1 was grown by wet chemistry \cite{Jardim} using the procedure described in ref. \cite{Munoz}, while
BMO2 was grown by a solid state reaction: stoichiometric amounts of Bi$_{2}$O$_{3}$ and Mn$_{2}$O$_{3}$ were thoroughly
mixed in an agate mortar, and heated to 800 $^{\circ}$C for 48 h and to 900 $^{\circ}$C for 72 h,
with intermediate grindings at each 24 h. For powder diffraction experiments, a sieve was used to
reject grains larger than $\sim 5$ $\mu$m, and the samples were deposited over flat Cu holders, apropriate for Bragg-Brentano
geometry. Synchrotron x-ray diffraction experiments in the single crystal and powder samples were performed on the XPD
beamline of the Laborat\'{o}rio Nacional de Luz S\'{i}ncrotron (LNLS), \cite{Ferreira} using an incident
beam with $\lambda =1.5499$ \AA\ for the single crystal study and $\lambda =1.3773$ \AA\ for the powder measurements,
except when otherwise stated. The beam was focused onto a spot $\sim 0.8 \times 2.0$ mm$^{2}$ for the single crystal and
$\sim 0.8 \times 5.0$ mm$^{2}$ for powders at the sample position.
A Ge(111) analyzer crystal was placed in a goniometer attached to the $2\theta$ arm, and a scintillation
detector was used. The instrumental resolution for this setup was $\sim 0.01 ^{\circ}$ full width at half maximum (FWHM) at
$2 \theta = 25 ^{\circ}$,\cite{Ferreira} and a step width of $0.0025 ^{\circ}$ was chosen for the powder measurements
while the sample was rocked by $1^{\circ}$ during each observation to minimize graininess effects.
For single crystal measurements, a natural (001) surface with no treatment was chosen, and the lattice parameters were obtained from
the Bragg positions of the (402), (045), and (006) reflections. Each reflection showed a clearly split three-peak structure in the axial
($\theta-2\theta$) scans at all temperatures, indicating the presence of at least three domains with distinct sets of
lattice parameters, possibly due to inhomogenously distributed Bi vacancies. All the domains showed
identical temperature-dependence for the lattice parameters, and the results shown in this work (including the peak widths)
are given for the strongest peak of each Bragg reflection.  
A closed cycle He cryostat was employed in our measurements, and the temperature was measured with an estimated accuracy better than $\sim 1$ K
and stability of $\sim 1$ mK. All the measurements were performed below $\sim 100$ K, while the base temperature (11 K for
powders and 17 K for the single crystal) was determined by the performance of the cryostat at the time of the experiments.
dc-magnetic susceptibility measurements were performed on a commercial superconduction quantum interference
device (SQuID) magnetometer, while the specific heat was measured on a commercial platform using the relaxation method.

\section{Results and Analysis}

\subsection{{\it dc}-magnetic susceptibility and Specific heat}

\begin{figure}
\includegraphics[width=0.9 \textwidth]{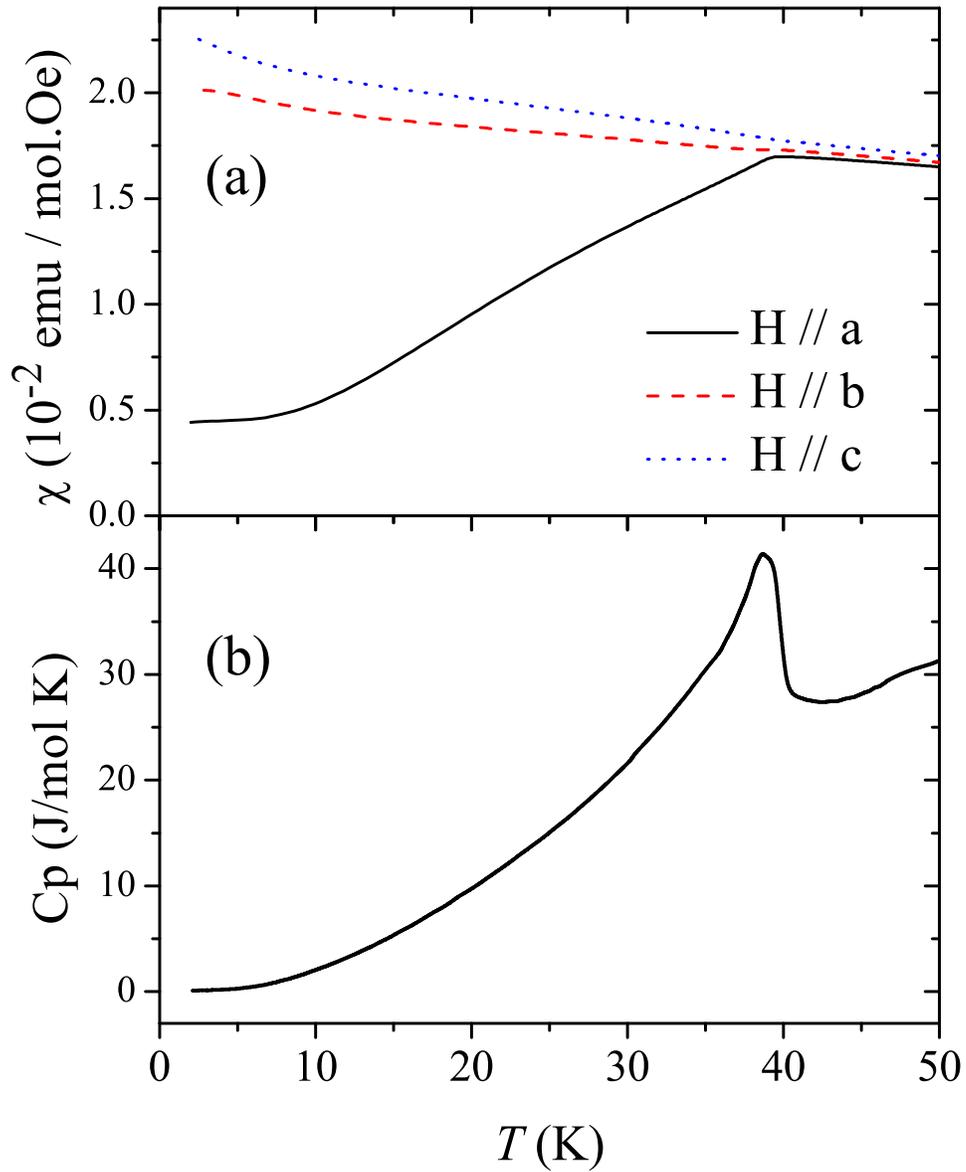}
\vspace{-1.0cm}
\caption{\label{bulk} (Color online) Temperature-dependence of (a) magnetic susceptibility taken with a magnetic
field of 2000 Oe along the {\bf a}, {\bf b}, and {\bf c} directions; and (b) specific heat of the BiMn$_{2}$O$_{5}$ single
crystal.}
\end{figure}

BiMn$_{2}$O$_{5}$ has been described as a non-colinear commensurate antiferromagnet at low temperatures with propagation vector
$\vec{\tau}=(\frac{1}{2},0,\frac{1}{2})$, and the spins pointing
nearly along the {\bf a}-direction, as inferred from neutron powder diffraction (NPD) measurements by Mu\~{n}oz {\it al.}\cite{Munoz}
As a preliminary bulk characterization,
we performed {\it dc}-magnetic susceptibility ($\chi_{dc}$) and specific heat ($C_{p}$) measurements on the single crystal
(see Figs. \ref{bulk}(a) and (b)). Only one transition could be unequivocally distinguished in our $C_{p}$ data, within
our resolution, while $\chi_{dc}$ data show that the Mn spins are indeed oriented nearly along the {\bf a}-direction in
the ordered phase,
as inferred by the smaller susceptibility below $T_{N}$ with the field along this direction. The $\chi_{dc}$ data are consistent
to those reported in Ref. \cite{Golovenchits}, and the conclusions drawn from $\chi_{dc}$ and $C_{p}$ data
are entirely consistent with the previous NPD results.\cite{Munoz} ($\chi_{dc}$) measurements on the two powder samples
also confirmed the AFM transition at $T_{N} \sim 38-40$ K.

\subsection{X-ray diffraction}

\begin{figure}
\includegraphics[width=0.9 \textwidth]{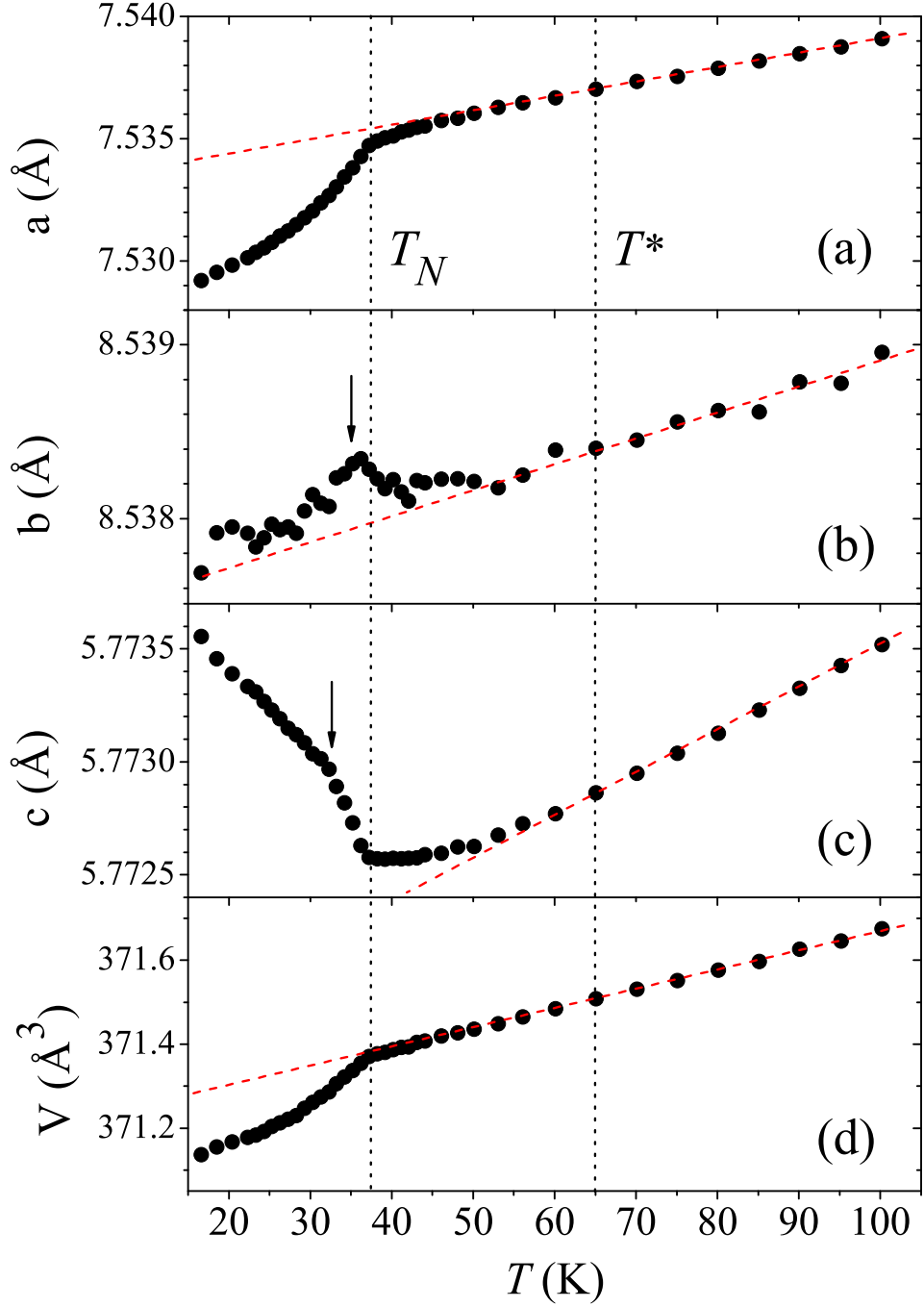}
\vspace{-1.0cm}
\caption{\label{lattpar} (Color online) Temperature-dependence of the {\it a}, {\it b}, and {\it c} lattice parameters,
and unit cell volume {\it V}. Dashed lines fitting the data above 65 K are guides to the eyes. The vertical lines mark
the magnetic ordering temperature $T_{N}$ and the onset of the strongly correlated paramagnetic state (see text and
ref. \cite{Ali}.}
\end{figure}

Figures \ref{lattpar}(a-c) show the temperature-dependence of the $a$, $b$, and $c$ lattice parameters.
The corresponding unit cell
volume $V$ is given in Fig. \ref{lattpar}(d). For temperatures between $\sim T^{*}=65$ K and 100 K (our upper limit in
this work), the evolution of all the lattice parameters follow a straight line within our resolution. The linear
thermal expansion coefficients in this $T$-range are: $\alpha_{a} = (1/a)\delta a/ \delta T=7.7(1) \times 10^{-6}$ K$^{-1}$;
$\alpha_{b}=1.7(2) \times 10^{-6}$ K$^{-1}$; $\alpha_{c}=3.27(4) \times 10^{-6}$ K$^{-1}$; and the volumetric expansion coefficient is
$\beta=1.27(3) \times 10^{-5}$ K$^{-1}$.
For $T_{N} < T \lesssim T^{*}$, deviations from this behavior were observed. Interestingly, such
deviations had a distinct sign for the $a$ lattice parameter with respect to $b$ and $c$, so that the unit cell volume
followed the constant thermal expansion down to $T_{N} \sim 38$ K. At $T_{N}$, clear anomalies were observed for all
lattice parameters and unit cell volume. While the $a$ lattice parameter show a contraction below
$T_{N}$, the $b$ parameter show a peak-like feature close to $T_{N}$ and $c$ shows an expansion on cooling. The dominant contraction
of $a$ shown in Fig. \ref{lattpar}(a) leads to a significant reduction of the unit cell volume below $T_{N}$ (see Fig. \ref{lattpar}(d)).

In order to correlate the observed lattice parameter anomalies to the antiferromagnetic order parameter and obtain more
detailed information on the magnetoelastic coupling in this material, the contributions from the non-magnetic thermal
expansion coefficient were subtracted, leading to
magnetoelastic contributions to $a$, $b$, and $c$, which we refer to as $a_{M}$, $b_{M}$, and $c_{M}$.
For this procedure, we assumed that the non-magnetic contributions to the lattice expansion have linear temperature-dependence
also below $T_{N}$, at least down to 17 K. These data were compared to the square of the
average sublattice magnetization of the Mn$^{4+}$ ($4f$ site) and Mn$^{3+}$ ($4h$ site) ions ($M^{2}$), extracted from ref. \cite{Munoz},
also normalized at 17 K (see Fig. \ref{magnetoelastic}). It is interesting to note that the evolution of
$a_{M}$ closely follows $M^{2}$ below $T_{N}$. The peak-like feature in the $b_{M}$ parameter takes place at $\sim 36$ K,
clearly below $T_{N}$, while $c_{M}$ appears to show a feature at $\sim 33$ K, increasing linearly on further cooling.
These anomalies for $b_{M}$ and $c_{M}$ below $T_{N}$ are indicated by arrows in Fig. \ref{lattpar}.

\begin{figure}
\includegraphics[width=0.9 \textwidth]{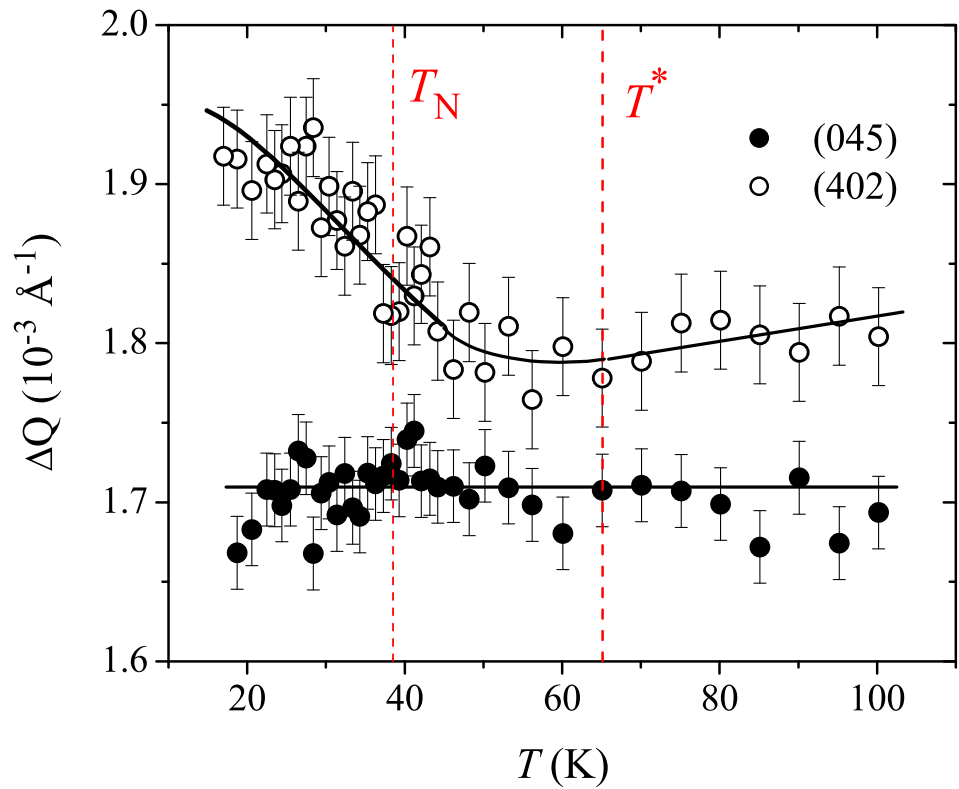}
\vspace{-1.0cm}
\caption{\label{width} (Color online) Bragg peak widths obtained from radial ($\theta-2\theta$) scans around the (402) and (045) reflections
for the single crystal. The solid lines are guides to the eyes.}
\end{figure}

Figure \ref{width} shows the width of the (402) and (045) reflections, revealing a broadening of (402) on cooling below $T^{*}$
and a nearly constant width of (045). Such anisotropic strain broadening indicates a magnetically-driven
fluctuation of the $a$ lattice parameter throughout the sample, in contrast to $b$ and $c$,
within our resolution. 

\begin{figure}
\includegraphics[width=0.9 \textwidth]{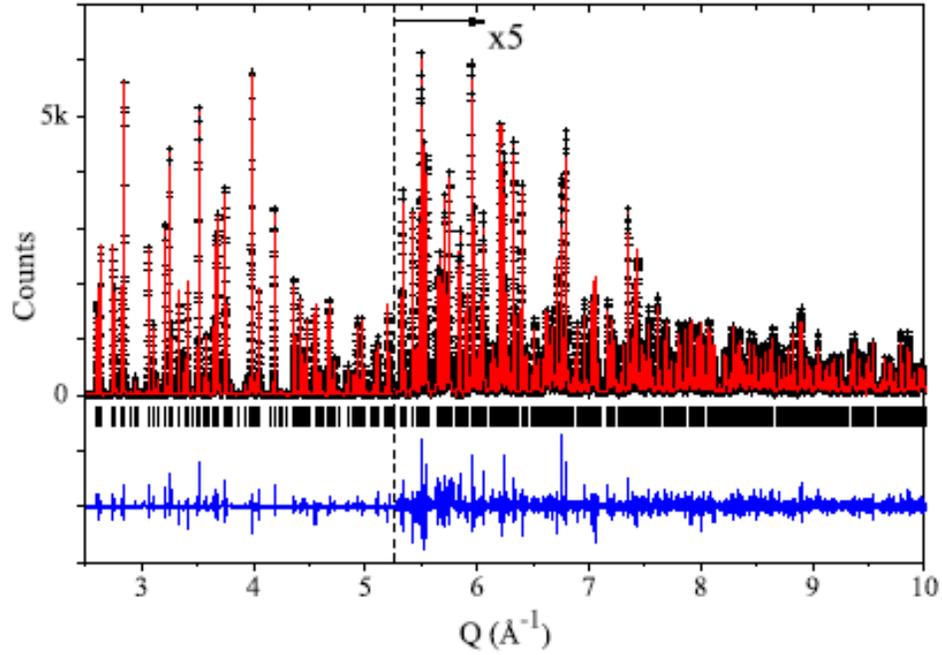}
\vspace{-1.0cm}
\caption{\label{profile} (Color online) (a) Observed (symbols), calculated (line) and difference (botton line)
x-ray powder diffraction profile of sample BMO2 at 100 K, taken with $\lambda = 1.1271$ \AA.
The goodness-of-fit factors are, $R_{p}=13.6$ \%\, and $R_{p}=24.9$ \%\ (background subtracted), and $\chi^{2}=1.85$.
The data were multiplied by five above $Q=5.3$ \AA$^{-1}$ for better visualization. The experimental data with $Q < 2.5$ \AA$^{-1}$ were
excluded from the refinement (see text).}
\end{figure}

Figure \ref{profile}(a) show the full powder diffraction pattern of BMO2 at 100 K. The crystal structure
was refined under the $Pbam$ space group using the GSAS+EXPGUI suite.\cite{gsas,expgui} 
The structure reported in Ref. \cite{Munoz} at 300 K was used as the initial model for the refinement. The calculated profile after the
refinement is also shown in Fig. \ref{profile}.
The experimental data with $Q < 2.5$ \AA$^{-1}$ were excluded from this refinement to avoid the instrumental or extrinsic
effects of peak asymmetry, self-absorption due to surface roughness and beam footprint larger than the sample size, which
become noticeable at lower angles for the reflection geometry employed here. This procedure does not affect
significantly the accuracy and precision of the refined parameters, since the density of Bragg peaks is larger in the higher $Q$-region.
The overall fitting quality is satisfactory, and the refined structural parameters
are given in Table \ref{parameters}. No impurity phases were observed within our
sensitivity. The equivalent pattern of BMO1 (not shown) revealed weak unidentified impurity peaks
($< 0.5$ \% of the strongest peak of the main phase. Overall, both samples
were found to be homogeneous and of very good crystalline quality above $T_{N}$.
No sign of anisotropic strain or symmetry lowering were observed for both samples at 100 K.

\begin{figure}
\includegraphics[width=0.9 \textwidth]{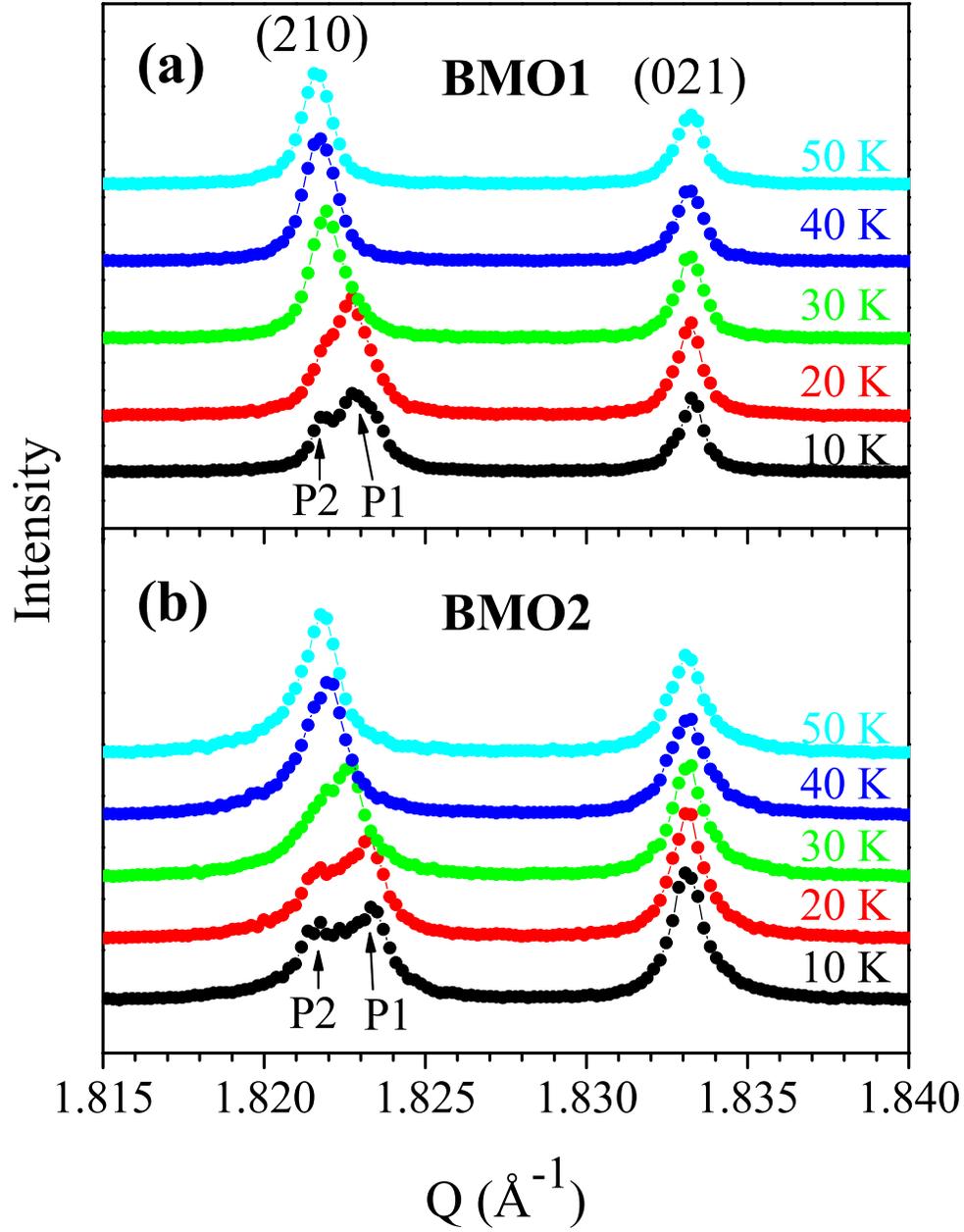}
\vspace{-1.0cm}
\caption{\label{split} (Color online) (a) Small portion of the powder diffraction profile of BMO1 and BMO2 samples
covering the (201) and (021) reflections at selected temperatures. The profiles were vertically translated for better
visualization.}
\end{figure}

Figures \ref{split}(a) and \ref{split}(b) show a selected portion of the powder diffraction profiles for BMO1 and BMO2, respectively,
including the (210) and (021) Bragg 
peaks, at several temperatures. It can be observed that, while the (021) peak shows nearly
no $T$-dependence, the (210) reflection splits in two peaks at low $T$. A similar splitting was clearly identified
in many other $(hkl)$ reflections with a large $h/(k+l)$ ratio. Attempts to index all Bragg peaks at 10 K within a single
crystallographic phase with either a monoclinic or triclinic unit cell derived from the high-temperature orthorhombic cell
were unsuccessfull. On the other hand, a model with two distinct phases with $Pbam$ symmetry and slightly different sets
of lattice parameters could match the observed Bragg peak splittings in the powder profile. These phases were
labeled as P1 and P2 in Figs. \ref{split}(a) and \ref{split}(b). In order to avoid
divergences in the fit, the $b$ lattice parameter had to be constrained as equal in both phases. The refined lattice
parameters for P1 and P2 at 10 K are given in Table \ref{P1P2}. It can be seen that P1 has smaller $a$
and slightly larger $c$ than P2. Unfortunately, the atomic parameters of P1 and P2 at 10 K could not be reliably extracted
from the refinement, since the fit did not converge when all the relevant atomic parameters were simultaneously refined.
This is mostly likely due to significant Bragg peak overlap of the two coexisting phases.

\begin{figure}
\includegraphics[width=0.9 \textwidth]{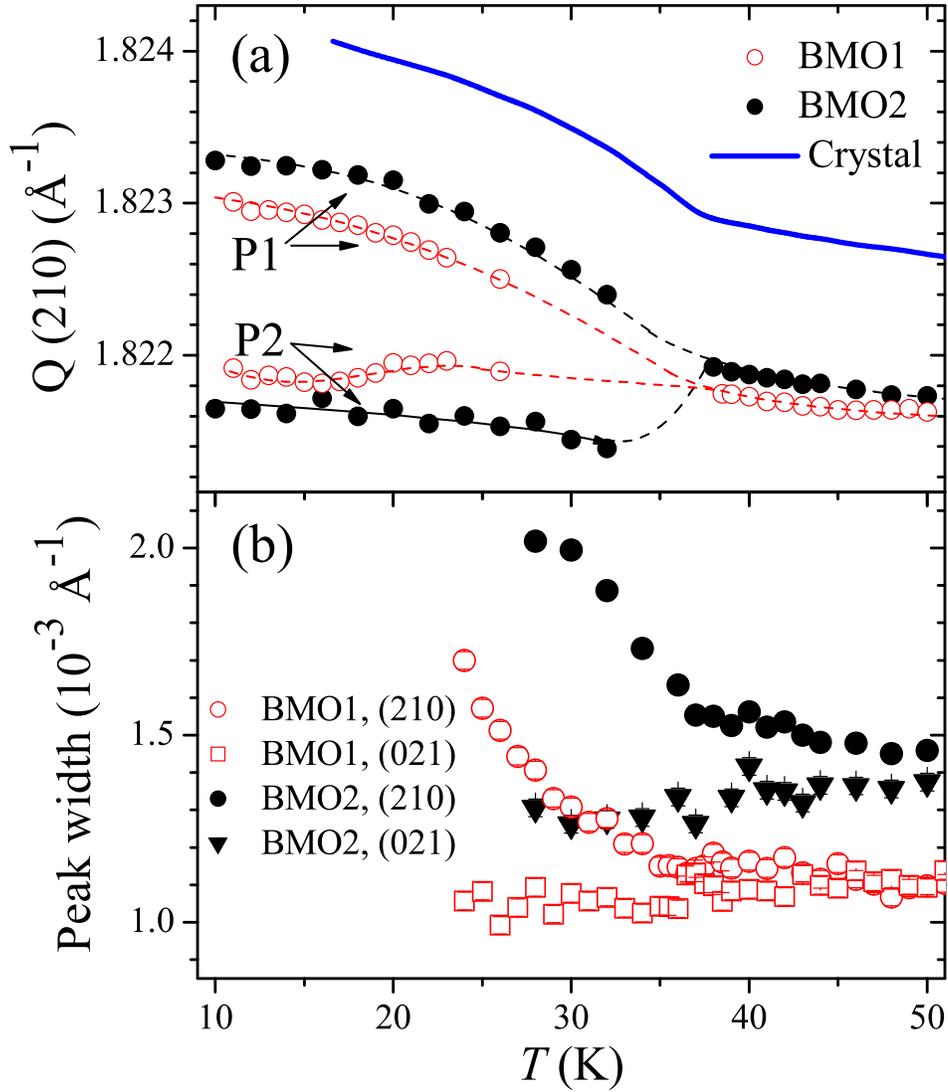}
\vspace{-1.0cm}
\caption{\label{powderresults} (Color online) (a) Symbols: temperature dependence of the positions of the (210) reflection
for P1 and P2 phases of BMO1 and BMO2 powder samples (see Fig. \ref{split}). The solid line indicates the expected behavior
taken from single crystal data shown in Fig. \ref{lattpar}. (b) Temperature dependence of the width of the (210) and (021)
reflections of BMO1 and BMO2, taken in the temperature region where the scattering at the (210) position could be fitted
by a single lorentzian peak.}
\end{figure}

To gain further insight into the nature of phases P1 and P2 below $T_{N}$, the Bragg peak positions of the (210) reflections
of both phases, obtained from a fit with using Lorentzian
lineshapes (symbols), are given in Fig. \ref{powderresults}(a) for samples BMO1 and BMO2.
For temperatures above $T_{N}$, the positions of the single (210) peak corresponding to the unique structural phase are given.
For the temperature interval near and below $T_{N}$ in which the two peaks could not be reliably separated in the fit,
the positions are not given in this figure. 
The ``expected'' positions of the (210) reflections, obtained from the single crystal
data of Figs. \ref{lattpar}(a-c), are given in Fig. \ref{powderresults}(a) (solid line). 
An analysis of the (210) and (021) peaks widths obtained with a single Lorentzian fit for each reflection,
is given in Fig. \ref{powderresults}(b). While the (021) width shows only a
weak temperature dependence, the (210) width increases steeply for temperatures below $\sim T_{N}$, for both
BMO1 and BMO2 samples. This result shows that the onset of structural phase coexistence on powder samples takes place near
the magnetic ordering temperature, $T_{N} \sim 38-40$ K. 

\section{Discussion}

We begin our discussion by considering the lattice parameter behavior above $T_{N}$ (see Fig. \ref{lattpar}). In this temperature region, the
thermal expansion is highly anisotropic, $\alpha_{a}$ being much larger than $\alpha_{b}$ and $\alpha_{c}$. Remarkably,
the unit cell volume $V$ shows a linear temperature dependence between $\sim T_{N}$ and 100 K within our resolution,
indicating a nearly constant thermal expansion coefficient $\beta$. This is a highly
unusual behavior, since $\beta = b_{3}T^{3}+b_{5}T^{5}+b_{7}T^{7}+...$ is expected at sufficiently low temperatures,\cite{Barron} and a constant
$\beta$ should occur only for $T>>\theta_{D}$, where $\theta_{D}=235$ K is the Debye temperature for
BiMn$_{2}$O$_{5}.$\cite{Munoz} A possible explanation for this intriguing behavior is the hypothetical presence of a
very low-frequency optical mode or some other dispersionless lattice excitation,
which thermal population might dominate the lattice expansion below 100 K. This possible mode or excitation might
be associated with coupled rigid rotations of the MnO$_{6}$ octahedra and MnO$_{5}$ pyramids, or most likely be
a rattling motion of Bi ions inside the BiO$_{8}$ cage. We should mention that
an anomalous distribution of Tb-O distances was observed for the related compound TbMn$_{2}$O$_{5}$, with relatively large
thermal dependence even at low temperatures,\cite{Tyson} also suggesting low-energy lattice excitations in this family.
Considering the linear volumetric thermal expansion above $T_{N}$, we expect that the energy of such mode should
satisfy the relation $E << k_{B}T_{N} =3.4$ meV. Preliminary Raman\cite{Ali2} and Infrared\cite{Massa} spectroscopy
measurements did not reveal any clear optical phonon with energies between $\sim 2$ and $\sim 5$ meV, indicating that the suggested lattice
excitation is either silent or have an energy below $\sim 2$ meV. We should mention that,
since the linear temperature dependence of $V$ occurs in the paramagnetic/paraelectric phase, it cannot
be explained by the possible presence of electromagnons in the ferroelectric/antiferromagnetic phase.\cite{Pimenov1,Sushkov,Aguilar,Pimenov2}

Even though the unit cell volume shows a linear temperature dependence down to $T_{N}$, the individual lattice parameters
$a$, $b$, and $c$ present deviations from a linear behavior below $T^{*} \sim 65$ K. The sign of this deviation is opposite
for $a$ than for $b$ and $c$ (see Fig. \ref{lattpar}(a-c)), compensating each other in the volumetric expansion between
$T_{N}$ and $T^{*}$. Also, these relatively small deviations from the linear temperature dependence
in the paramagnetic region have the same sign of the much larger magnetoelastic anomalies below $T_{N}$, i.e., a contraction for
$a$ and an expansion for $b$ and $c$ on cooling. This correspondence suggests that the deviations from linear temperature
dependence between $T_{N}$ and $T^{*}$ are not driven by phonons,
but are rather related to the strong short-range spin correlations in this temperature region,
consistent with a previous Raman scattering study in BiMn$_{2}$O$_{5}$.\cite{Ali} The magnetically-driven
strain broadening of the (402) reflection below $\sim T^{*}$ (see Fig. \ref{width}) is an additional evidence
that short-range spin correlations may influence the crystal lattice of BiMn$_{2}$O$_{5}$.

\begin{figure}
\includegraphics[width=0.9 \textwidth]{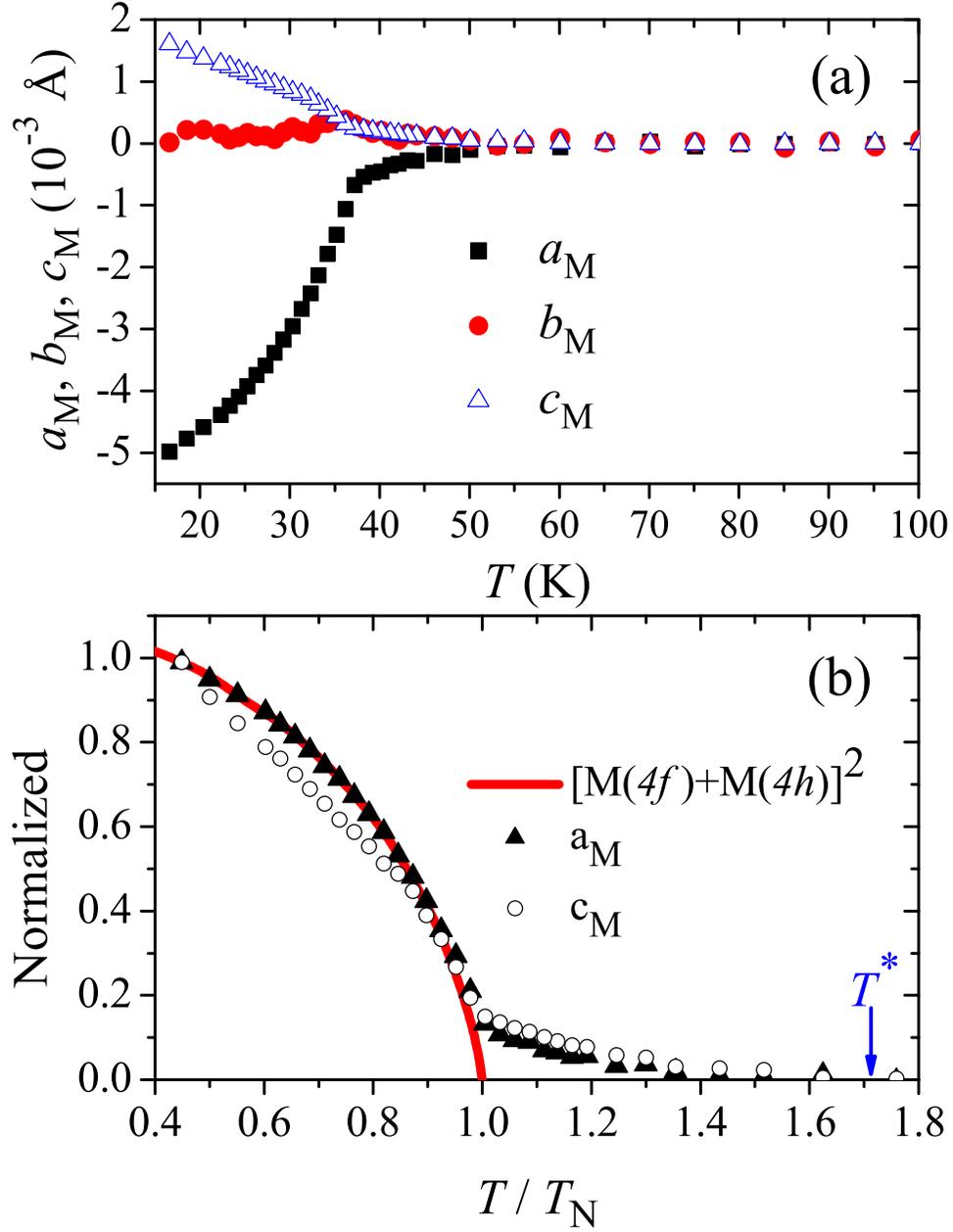}
\vspace{-1.0cm}
\caption{\label{magnetoelastic} (Color online) (a) Magnetoelastic component of $a$, $b$, and $c$ lattice parameters,
defined as the difference between the symbols and straight line in Fig. \ref{lattpar} and labeled as $a_{M}$, $b_{M}$, and $c_{M}$,
respectively. (b) Scaling of $a_{M}$ and $c_{M}$, normalized at 17 K (symbols), with the
squared average sublattice magnetization of Mn$^{4+}$ and Mn$^{3+}$ ions (solid line),
extracted from ref. \cite{Munoz}.}
\end{figure}

The most noticeable lattice anomalies takes place below $T_{N}$, deserving a careful consideration.
In a quadratic spin hamiltonian for transition-metal compounds, 
the magnetic energy and the atomic displacements due to exchange striction are
proportional to the square of the magnetic order parameter ($M^{2}$), in a mean-field approximation, assuming that the angle
between ordered spins remains constant below $T_{N}$. This proposition is valid for either Heisenberg, Ising, or
Dzyaloshinskii-Moriya spin hamiltonians, or a combination of them.
In order to accomodate the atomic movements directly related to
exchange striction, the crystal lattice relaxes, possibly leading to complex structural changes even for atoms not directly
related to the exchange mechanism. For small displacements, the magnitude of the elastic response of the lattice 
is directly proportional to the perturbing displacements due to exchange striction, therefore the overal
lattice anomalies arising from exchange striction are expected to follow $M^{2}$ for a quadratic spin hamiltonian.
Figure \ref{magnetoelastic}(b) shows 
that this simple prediction is confirmed for $a_{M}$, but clearly fails for $b_{M}$ and $c_{M}$ in the whole studied temperature interval.
In addition, these two parameters
show interesting features at $\sim 36$ K and $\sim 33$ K that might be associated to additional magnetic transitions
below $T_{N}$. We especulate that these anomalies might be due to spin-flip
and/or incommensurate-commensurate magnetic transitions, such as observed in other members of the family.\cite{delaCruz}
We should mention that no evidence of multiple magnetic transitions has been observed in our $C_{p}$ data, within our temperature
resolution, while a two-peak structure in $C_{p}$ close to $T_{N}$ was previously reported.\cite{Munoz}
More detailed neutron diffraction experiments in the temperature interval close to $T_{N}$ may be necessary to confirm or dismiss
this hypothesis.

The fact that only $a_{M}$ scales with $M^{2}$ is interesting. This is
also the direction in which the magnitude of the lattice anomaly is the largest.
This can be rationalized on the basis of the magnetic
structure of BiMn$_{2}$O$_{5}$ given in Ref. \cite{Munoz}. According to this, along the {\bf a}-direction
all the nearest-neighbor spin alignment between Mn$^{4+}$O$_{6}$ octahedra and Mn$^{3+}$O$_{5}$ pyramids and between
consecutive Mn$^{3+}$O$_{5}$ pyramids in the sequence ...Mn$^{4+}$-Mn$^{3+}$-Mn$^{3+}$-Mn$^{4+}$...
are nearly AFM, with no clear manifestation of magnetic frustration except perhaps for a relatively small non-colinearity of the
spin aligment. On the other hand, the spin alignment along {\bf b} and {\bf c} directions show clear signs of 
competing interactions. Along {\bf b}, the coupling between Mn$^{4+}$O$_{6}$ octahedra and
Mn$^{3+}$O$_{5}$ pyramids alternates between FM and AFM, breaking the inversion symmetry of the structure and possibly
causing ferroelectricity as in other members of the $R$Mn$_{2}$O$_{5}$ family.\cite{Cheong} Along {\bf c}, the spin aligment
between edge-shared Mn$^{4+}$O$_{6}$ octahedra also alternates between FM and AFM, the AFM and FM alignments corresponding to the
shorter (=2.767 \AA) and longer (=2.988 \AA) Mn$^{4+}$-Mn$^{4+}$ distances, respectively. Bi and Mn$^{3+}$ planes are intercalated
between the AFM- and FM-coupled Mn$^{4+}$ planes, respectively. It has been argued that the Mn$^{4+}$-O-Mn$^{3+}$-O-Mn$^{4+}$
superexchange path and the longer Mn$^{4+}$-Mn$^{4+}$ distances may help stabilizing the FM coupling among half of the
Mn$^{4+}$ pairs.\cite{Munoz}
We mention that, while the Mn spin structures in the {\it ab} plane are similar for
all members of the family, the alignment along {\bf c} is strongly dependent on the $R$-ions, with the component of
the spin propagation vector varying from $k_{z}=1/4$ for $R=$Er\cite{Buisson} to $k_{z}=1/2$ for $R=$Bi.\cite{Bertaut,Munoz}
This is another clear indication of competing magnetic interactions along this direction.

The distinct features of
the magnetic ordering along each direction obviously leads to an anisotropic magnetoelastic coupling, as manifested in
the data of Fig. \ref{lattpar}. First of all, the AFM alignment of nearest-neighbor spins along {\bf a} leads to a relatively large
contraction of the lattice along this direction below $T_{N}$ (see Fig. \ref{lattpar}(a)), in order to enhance the
AFM coupling of the ...Mn$^{4+}$-Mn$^{3+}$-Mn$^{3+}$-Mn$^{4+}$... chains. On the other hand, some of the 
magnetic interactions are frustrated along {\bf b}, and therefore the atomic distances related to `satisfied'
pairs are expected to be reduced and others (between `frustrated' pairs) should increase below $T_{N}$, leading to a compensation
which is manifested as small values of $b_{M}$ in comparison to $a_{M}$ (see Fig. \ref{magnetoelastic}).
In the case of $c_{M}$, a presumed increase of the separation
of FM Mn$^{4+}$ pairs may be partially compensated by a corresponding approximation between the AFM Mn$^{4+}$ pairs or vice-versa,
leading to smaller $c_{M}$ anomalies in comparison to $a_{M}$.
In addition, the evolution of $b_{M}$ and $c_{M}$ may be also influenced by a lattice relaxation in response to the
relatively large contraction of $a_{M}$. Even considering such a complex situation, it is expected that magnetoelastic
anomalies along all directions should follow the square of the sublattice magnetization for a spin hamiltonian composed
only of quadratic terms on spin, as argued above. The fact that $c$ and most notably $b$ do not follow this behavior even for
the temperature interval where no additional phase transitions are evidenced ($T < 33$ K) suggest that
other terms, possibly of fourth order (biquadratic, three-spin and four-spin),\cite{Kobler} should be included in the
spin hamiltonian to correctly account for the magnetoelastic anomalies in BiMn$_{2}$O$_{5}$. Although these terms are
expected to be significantly smaller than the quadratic exchange terms, the magnetic frustration caused by the complex lattice
geometry may cause a nearly complete cancellation of the quadratic terms, leading to a relative increase of importance
of the quartic terms, and consequently a manifestation of such terms in the magnetoelastic anomalies in the directions
where the spin frustration is most pronounced. This conclusion is likely extensible to the other members of the 
$R$Mn$_{2}$O$_{5}$ family and may be of relevance to a quantitative understanding of their complex magnetic phase diagram
and multiferroic properties.

The analysis of the powder samples provides relevant additional information. Contrary to the single crystal,
a coexistence of two distinct phases with orthorhombic metrics (within our resolution) and slightly different sets of
$a$ and $c$ lattice parameters was observed below $\sim T_{N}$ for both BMO1 and BMO2 powder samples. The lattice parameters of P1 
are closer to P2 in BMO1 than in BMO2, indicating that the magnitude of this effect in powder samples is dependent on the
sample details. The phase fractions of P1 and P2 are also strongly sample-dependent (see Table \ref{P1P2}). 
Also, a comparison of the $T$-dependence of the position of a particular Bragg peak with temperature
of both powder samples with that expected from single crystal data (see Fig. \ref{powderresults}) shows that the phase labeled
P1 (with smaller $a$ lattice parameter) corresponds to the magnetostructural ground state of BiMn$_{2}$O$_{5}$, while P2
may be either a metastable state or a phase stabilized by slight chemical inhomogeneities (see below).
It is well known that powder samples may favor the presence of metastable
structural phases not generally observed on single crystals, leading to polymorphism.
In the present case, the structures of P1 and P2 converge above $T_{N}$, showing that the phase separation
cannot be trivially explained by large chemical inhomogeneities between distinct grains. In fact,
P2 presumably shows a different magnetic structure than P1, leading to distinct
magnetoelastic coupling effects and therefore a slightly different set of lattice parameters at low temperatures. On the
other hand, slight chemical inhomogeneities (such as in the Bi occupancy) not clearly evidenced in our powder diffraction
profiles above $T_{N}$ might still play a role in stabilizing a magnetically phase-separated state below $T_{N}$ if
there is a pre-existing close competition between distinct possible magnetic ground states. The fact
that P1 and P2 unit cells show differences in $a$ and $c$ lattice parameter with no observable change in $b$
suggests that magnetic structures of these phases show distinct components of the propagation vector along {\bf a} and/or {\bf c}.
Further studies are necessary to confirm or dismiss this hypothesis.

\section{Conclusions}

In summary, the magnetoelastic anomalies in single crystal and powdered BiMn$_{2}$O$_{5}$ were investigated in detail by
high-resolution synchrotron x-ray diffraction.
It was found in the single crystal study that this compound shows a linear temperature dependence of the unit cell volume
between $T_{N}$ and 100 K, well below the Debye temperature,
which has been associated to the possible presence of a low-energy lattice excitation that still remains to be directly observed.
Below $T_{N}$, significant lattice parameter anomalies were found,
most notably in the $a$-parameter, due to a magnetoelastic coupling. The anomaly of $a$ was found to follow the square
of the average sublattice magnetization, as expected for a quadratic spin hamiltonian. The anomalies of $b$ and $c$ follow
a more complex behavior, signaling the existence of additional transitions below $T_{N}$ as in other members of the
$R$Mn$_{2}$O$_{5}$ family, and perhaps higher-order terms in the spin hamiltonian.
It was argued that strong short-range spin correlations between $T_{N}$ and $T^{*}=65$ K\cite{Ali} give rise
to an observable contribution to the linear thermal expansion coefficients. X-ray powder diffraction measurements taken on samples grown
by distinct routes show phase coexistence of different magnetostructural states below $T_{N}$. One of these states
corresponds to that observed in the single crystal, while the other was attributed to a competing magnetic state which
might be metastable or stabilized by subtle chemical inhomogeneities that might be present in the powder samples.

\section{ACKNOWLEDGEMENTS}

We thank L.C.M. Walmsley, N.E. Massa, and R.F. Jardim for helpful discussions.
This work was supported by Fapesp and CNPq, Brazil, Russian Foundation for Basic Research, Presidium of RAS, and Division of Physics of RAS.

\newpage

\begingroup
\begin{table*}
\caption{\label{parameters} Refined lattice and atomic parameters of sample BMO2 at 100 K.
Errors in parentheses are statistical only, and represent one standard deviation.}
\begin{ruledtabular}
\begin{tabular}{c c c c c}
$T = 100$ K & Pbam & $a=7.54116(1)$ \AA & $b=8.52994(1)$ \AA & $c=5.75437(1)$ \AA \\
\end{tabular}
\begin{tabular}{c c c c c c c}
Atom & site & $x$ & $y$ & $z$ & $U_{iso}$ (\AA$^{2})$ & frac \\
Bi & $4g$ & 0.15896(4) & 0.16556(4) & 0 & 0.00588(6) & 0.938(4) \\
Mn1 & $4f$ & 1/2 & 0 & 0.2596(2) & 0.0021(2) & 1 \\
Mn2 & $4h$ & 0.40755(15) & 0.35091(14) & 1/2 & .0029(2) & 1 \\
O1 & $4e$ & 0 & 0 & 0.2876(10) & 0.0048(5) & 1 \\
O2 & $4g$ & 0.1567(8) & 0.4453(6) & 0 & 0.0048(5) & 1 \\
O3 & $4h$ & 0.1437(7) & 0.4243(6) & 1/2 & 0.0048(5) & 1 \\
O4 & $8i$ & 0.3866(5) & 0.2018(4) & 0.2525(7) & 0.0048(5) & 1 \\
\end{tabular}
\begin{tabular}{c c c}
$R_{p}=13.6$ \% & $R_{wp}=25.3$ \% & $\chi^{2}=1.86$ \\
\end{tabular}
\end{ruledtabular}
\end{table*}
\endgroup

\begingroup
\begin{table*}
\caption{\label{distances} Interatomic distances for sample BMO2 at 100 K (\AA). Errors in parentheses are statistical
only, and represent one standard deviation.}
\begin{ruledtabular}
\begin{tabular}{c c c c c c c}
& Mn$^{4+}$O$_{6}$ & & & & Mn$^{3+}$O$_{5}$ & \\
\end{tabular}
\begin{tabular}{c c c c}
Mn1-O2$(\times 2)$ & 1.961(4) & Mn2-O1$(\times 2)$ & 1.897(4) \\
Mn1-O3$(\times 2)$ & 1.872(4) & Mn2-O3$(\times 1)$ & 2.086(6) \\
Mn1-O4$(\times 2)$ & 1.923(4) & Mn2-O4$(\times 2)$ & 1.916(4) \\
$<$Mn1-O$>$ & 1.919(2) & $<$Mn2-O$>$ & 1.966(3) \\
\end{tabular}
\begin{tabular}{c c c c c c c}
& & & BiO$_{8}$ & & & \\
\end{tabular}
\begin{tabular}{c c c c}
Bi-O1$(\times 2)$ & 2.484(4) & & \\
Bi-O2$(\times 1)$ & 2.337(5) & Bi-O2$(\times 1)$ & 2.386(5) \\
Bi-O4$(\times 2)$ & 2.270(4) & Bi-O4$(\times 2)$ & 2.758(4) \\
$<$Bi-O$>$ & 2.468(2) & & \\
\end{tabular}
\end{ruledtabular}
\end{table*}
\endgroup

\begingroup
\begin{table*}
\caption{\label{P1P2} Lattice parameters and unit cell volume for phases P1 and P2 of powder samples BMO1 and BMO2 at 10 K.
Errors in parentheses are statistical only, and represent one standard deviation.}
\begin{ruledtabular}
\begin{tabular}{c c c c c c c}
Sample & Phase & fraction (\%) & $a$(\AA) & $b$(\AA) & $c$(\AA) & $V$(\AA$^{3}$ \\
BMO1 & P1 & 90 & 7.53039(3) & 8.52766(3) & 5.75426(3) & 369.517(3) \\
BMO1 & P2 & 10 & 7.53730(7) & 8.52766(3) & 5.75377(10) & 369.827(6) \\
BMO2 & P1 & 53 & 7.53094(3) & 8.52955(3) & 5.75497(3) & 369.673(3) \\
BMO2 & P2 & 47 & 7.53888(4) & 8.52955(3) & 5.75405(3) & 370.004(3) \\
\end{tabular}
\end{ruledtabular}
\end{table*}
\endgroup

\end{document}